\documentclass[twocolumn]{aastex631}
\usepackage{bm}   

\newcommand{\nc}{\newcommand*} 
\nc{\al}{\alpha}
\nc{\dt}{\delta}
\nc{\Gm}{\Gamma}
\def\({\left(}
\def\){\right)}
\def\[{\left[}
\def\]{\right]}
\def\e{\begin{equation}}
\def\q{\end{equation}}
\def\m{\begin{eqnarray}}
\def\n{\end{eqnarray}}
\nc{\Eq}[1]{Eq.~(\ref{#1})}     
\nc{\Fig}[1]{Fig.~\ref{#1}}     
\nc{\Table}[1]{Table~\ref{#1}}  
\nc{\Sec}[1]{Sec.~\ref{#1}}     
\nc{\Msun}{M_\odot}             
\nc{\ogw}{\Omega_{\mathrm{GW}}}
\nc{\km}{\mathrm{km}}
\nc{\Mpc}{\mathrm{Mpc}}
\nc{\fyr}{f_{\mathrm{yr}}}
\nc{\addref}{[\textcolor{red}{add ref}] } 
\nc{\eg}{\textit{e.g.~}}
\nc{\app}{\approx}
\nc{\hf}{\frac{1}{2}}
\nc{\discuss}{\textcolor{red}{Add discussion here!}}
\nc{\red}[1]{\textcolor{red}{#1}}
\nc{\A}[1]{\mathcal{A}_{#1}}
\nc{\Ogw}[1]{\Omega_{\mathrm{#1}}}
\nc{\bn}[1]{\dt\bm{t}_{\text{#1}}}
\nc{\bC}[1]{\bm{C}_{\text{#1}}}
\nc{\NTOA}{N_{\text{TOA}}}
\nc{\Nmode}{{N_{\text{mode}}}}
\nc{\ARN}{A_{\rm{SN}}}
\nc{\gRN}{\gamma_{\rm{SN}}}
\nc{\ADM}{A_{\rm{DM}}}
\nc{\gDM}{\gamma_{\rm{DM}}}
\nc{\bS}{\mathbf{\Sigma}}
\nc{\br}{\mathbf{r}}
\nc{\bN}{\mathbf{R}}
\nc{\Agw}{A_\mathrm{GWB}}
\nc{\UCP}{\mathrm{UCP}}
\nc{\TT}{\mathrm{TT}}
\nc{\ST}{\mathrm{ST}}
\nc{\SL}{\mathrm{SL}}
\nc{\VL}{\mathrm{VL}}
\nc{\BF}{\mathcal{BF}}

\begin{document}
\received{}
\revised{}
\accepted{}
\published{}
	
\title{Searching for Isotropic Stochastic Gravitational-Wave Background in the International Pulsar Timing Array Second Data Release}

\author{Zu-Cheng Chen}
\email{chenzucheng@itp.ac.cn} 
\affiliation{CAS Key Laboratory of Theoretical Physics, Institute of Theoretical Physics, Chinese Academy of Sciences, Beijing 100190, China}
\affiliation{School of Physical Sciences, University of Chinese Academy of Sciences, No. 19A Yuquan Road, Beijing 100049, China}
	
\author{Yu-Mei Wu}
\email{wuyumei@itp.ac.cn} 
\affiliation{CAS Key Laboratory of Theoretical Physics, Institute of Theoretical Physics, Chinese Academy of Sciences, Beijing 100190, China}
\affiliation{School of Physical Sciences, University of Chinese Academy of Sciences, No. 19A Yuquan Road, Beijing 100049, China}

\author{Qing-Guo Huang}
\email{Corresponding author: huangqg@itp.ac.cn}
\affiliation{CAS Key Laboratory of Theoretical Physics, 
    Institute of Theoretical Physics, Chinese Academy of Sciences,
    Beijing 100190, China}
\affiliation{School of Physical Sciences, 
    University of Chinese Academy of Sciences, 
    No. 19A Yuquan Road, Beijing 100049, China}
\affiliation{School of Fundamental Physics and Mathematical Sciences,
    Hangzhou Institute for Advanced Study, UCAS, Hangzhou 310024, China}
\affiliation{Center for Gravitation and Cosmology, 
    College of Physical Science and Technology, 
    Yangzhou University, Yangzhou 225009, China}

\date{\today}

\begin{abstract}
We search for isotropic stochastic gravitational-wave background (SGWB) in the International Pulsar Timing Array second data release. By modeling the SGWB as a power-law, we find very strong Bayesian evidence for a common-spectrum process, and further this process has scalar transverse (ST) correlations allowed in general metric theory of gravity as the Bayes factor in favor of the ST-correlated process versus the spatially uncorrelated common-spectrum process is $30\pm 2$. The median and the $90\%$ equal-tail amplitudes of ST mode are $\mathcal{A}_{\mathrm{ST}}= 1.29^{+0.51}_{-0.44} \times 10^{-15}$, or equivalently the energy density parameter per logarithm frequency is $\Omega_{\mathrm{GW}}^{\mathrm{ST}} = 2.31^{+2.19}_{-1.30} \times 10^{-9}$, at frequency of 1/year. However, we do not find any statistically significant evidence for the tensor transverse (TT) mode and then place the $95\%$ upper limits as $\A{\TT}< 3.95 \times 10^{-15}$, or equivalently $\ogw^\TT < 2.16 \times 10^{-9}$, at frequency of 1/year.
\end{abstract}

\section{Introduction} 

After the direct detection of gravitational waves (GWs) from a binary black hole \citep{LIGOScientific:2016aoc} and a binary neutron star \citep{LIGOScientific:2017vwq} mergers by LIGO-Virgo, other types of GW sources are yet to be identified. Of particular interest is the stochastic gravitational-wave background (SGWB) produced by the superposition of a large number of independent GW signals from compact binary coalescences. While the ground-based interferometers are sensitive to GWs from Hz to kHz, a pulsar timing array  \cite[PTA;][]{1978SvA....22...36S,Detweiler:1979wn,1990ApJ...361..300F}, which regularly monitors the time of arrivals (TOAs) of radio pulses from an array of stable millisecond pulsars, offers a unique and powerful probe to correlated signals at low frequencies from nHz to $\mu$Hz. The SGWB sources for PTAs could come from the inspiral of supermassive black hole binaries (SMBHBs) \citep{Jaffe:2002rt,Sesana:2008mz,Sesana:2008xk}, the first-order phase transition \citep{Witten:1984rs,Hogan:1986qda}, and the scalar-induced GWs \citep{Saito:2008jc,Yuan:2019udt,Yuan:2019wwo}, etc. It is expected that the SGWB from SMBHBs will be the first GW signal to be detected with PTAs \citep{Rosado:2015epa}.

There are three major PTAs with accumulated pulsar-timing data of more than a decade, namely the European Pulsar Timing Array \citep[EPTA;][]{Kramer:2013kea}, the North American Nanoherz Observatory for Gravitational Waves \citep[NANOGrav;][]{McLaughlin:2013ira}, and the Parkes Pulsar Timing Array \citep[PPTA;][]{Manchester:2012za}. These collaborations support the International Pulsar Timing Array \citep[IPTA;][]{Hobbs:2009yy,Manchester:2013ndt}. Over the last decade, PTAs have accumulated increasingly sensitive data sets, and the null-detection of GWs with PTAs has successfully constrained various astrophysical scenarios, such as cosmic strings \citep{Lentati:2015qwp,Arzoumanian:2018saf,Yonemaru:2020bmr}, SGWBs from SMBHBs with power-law spectra \citep{Lentati:2015qwp,Shannon:2015ect,Arzoumanian:2018saf}, and primordial black holes \citep{Chen:2019xse}, etc. It is widely expected that the inkling of an SGWB will first manifest as the emergence of a spatially uncorrelated common-spectrum process (UCP) among all pulsars, and culminate in the appearance of the spatial correlations that unambiguously signify the detection of an SGWB.
 
In a recent analysis, the NANOGrav collaboration found strong evidence for a stochastic common-spectrum process modeled by a power-law spectrum in their 12.5-yr data set \citep{Arzoumanian:2020vkk}. However, there was no statistically significant evidence for the tensor transverse (TT) spatial correlations which are deemed to be necessary to claim an SGWB detection consistent with general relativity. \cite{Chen:2021wdo} then reanalyzed the NANOGrav 12.5-yr data set and found strong Bayesian evidence that the common-spectrum process reported by NANOGrav collaboration has the scalar transverse (ST) spatial correlations which can originate from a general metric theory of gravity. Later on, the PPTA collaboration analyzed their second data release (DR2) and also found a common-spectrum process but without significant evidence for, or against, the TT spatial correlations \citep{Goncharov:2021oub}. Furthermore, \cite{Wu:2021kmd} searched for the non-tensorial polarizations in the PPTA DR2, and found no significant evidence supporting the existence of alternative polarizations, thus constraining the amplitude of each polarization mode. Note that the $95\%$ upper limit on the amplitude of the ST mode from \cite{Wu:2021kmd} is consistent with the result from \cite{Chen:2021wdo}.

The IPTA collaboration has published their second data release comprising of 65 pulsars in total with a timespan as long as about three decades \citep{Perera:2019sca}. The IPTA DR2 is an invaluable complement to the NANOGrav 12.5-yr data set and PPTA DR2 in the sense of more pulsars included and longer timespan. In this letter, we aim to search for the SGWB signal modeled as a power-law spectrum in the IPTA DR2 with a particular interest in exploring if the ST spatial correlations are present in the data set or not. 

\section{The Data Set and Methodology}

The IPTA DR2 \citep{Perera:2019sca} was created by combining published data from individual PTA data releases, including EPTA DR1 \citep{Desvignes:2016yex}, NANOGrav 9-yr data set \citep{NANOGrav:2015qfw}, and PPTA DR1 \citep{Manchester:2012za}. It consists of 65 pulsars and provides a better sky coverage compared to the IPTA DR1 \citep{Verbiest:2016vem}. There are two data combination versions in the IPTA DR2, namely VersionA and VersionB. The two versions are different in modeling the dispersion measure (DM) variation and handling the noise properties of pulsars \citep{Perera:2019sca}. In particular, the noise parameters of the pulsars in VersionB are re-estimated based on the IPTA data combination, while VersionA uses the previously constrained values from other PTA data sets \citep{Perera:2019sca}. It has been shown that the overall time-dependent DM variations modeled by these two methods are largely consistent with each other \citep{Perera:2019sca}. In this work, we use the VersionB data release by choosing only pulsars with a timing baseline greater than three years in our analyses and excluding the pulsar J1939+2134 because of its complicated DM variation and timing noise \citep{Kaspi:1994hp,Manchester:2012za,Lentati:2016ygu}.  Therefore, all results in this work are based on 52 pulsars that meet the requirements.

\begin{table*}[htbp!]
    \centering
    \scriptsize
    \caption{Parameters and their prior distributions used in the analyses.}
    \label{tab:priors}
    \begin{tabular}{llll}
        \hline\hline
        parameter & description & prior & comments \\
        \hline
        \multicolumn{4}{c}{Red Noise} \\[1pt]
        $\ARN$ & red-noise power-law amplitude & log-Uniform $[-20, -11]$ & one parameter per pulsar  \\
        $\gRN$ & red-noise power-law spectral index & Uniform $[0, 7]$ & one parameter per pulsar \\
        \hline
        \multicolumn{4}{c}{DM Noise} \\[1pt]
        $\ADM$ & red-noise power-law amplitude & log-Uniform $[-20, -11]$ & one parameter per pulsar  \\
        $\gDM$ & red-noise power-law spectral index & Uniform $[0, 7]$ & one parameter per pulsar \\
        \hline
        \multicolumn{4}{c}{Annual DM variation}\,\\
        $\A{Y}$ & annual DM variation amplitude & log-Uniform $[-10, -2]$ & one parameter per annual event \\
        $\phi_{\mathrm{Y}}$ & annual DM variation phase & Uniform $[0, 2\pi]$ &one parameter per annual event \\
        \hline
        \multicolumn{4}{c}{DM exponential dip}\,\\
        $\A{E}$ & exponential dip amplitude & log-Uniform $[-10, -2]$ & one parameter for pulsar J1713+0747 \\
        $ t_{\mathrm{E}}[\mathrm{MJD}]$ &time of the event & Uniform $[54500, 55000]$ & one parameter for pulsar J1713+0747 \\
        $\tau_{\mathrm{E}}[\mathrm{MJD}]$ & relaxation time for the dip & log-Uniform $[0, 2.5]$ & one parameter for pulsar J1713+0747 \\
        \hline
        \multicolumn{4}{c}{White Noise} \\[1pt]
        $E_{k}$ & EFAC per backend/receiver system & Uniform $[0, 10]$ & single-pulsar analysis only \\
        $Q_{k}$[s] & EQUAD per backend/receiver system & log-Uniform $[-8.5, -5]$ & single-pulsar analysis only \\
        $J_{k}$[s] & ECORR per backend/receiver system & log-Uniform $[-8.5, -5]$ & single-pulsar analysis only \\
        \hline
        \multicolumn{4}{c}{Common-spectrum Process} \\[1pt]
        $\A{\UCP}$ & UCP power-law amplitude & log-Uniform $[-18, -14]$ & one parameter for PTA \\
        & & log-Uniform $[-18, -11]$ ($\gamma_{\UCP}$ varied) & one parameter for PTA \\
        $\gamma_{\UCP}$ & UCP power-law spectral index & delta function ($\gamma_{\UCP}=13/3$) & fixed \\
        & & Uniform $[0, 7]$ ($\gamma_{\UCP}$ varied) & one parameter for PTA \\
        $\A{\TT}$ & GW amplitude of TT polarization & log-Uniform $[-18, -14]$ & one parameter for PTA \\
        $\A{\ST}$ & GW amplitude of ST polarization & log-Uniform $[-18, -14]$ & one parameter for PTA \\
        $\al$ & parameter in \Eq{alpha} & Uniform $[-10, 10]$ & one parameter for PTA \\
        \hline
    \end{tabular}
\end{table*}

The GWs will manifest as the unexplained residuals in the pulsar TOAs after subtracting a deterministic timing model that accounts for the pulsar spin behavior and the geometric effects due to the motion of the pulsar and the Earth \citep{1978SvA....22...36S,Detweiler:1979wn}. For two pulsars $a$ and $b$, the cross-power spectral density of the timing residuals induced by an SGWB at frequency $f$ is \citep{2008ApJ...685.1304L,Chamberlin:2011ev,Gair:2015hra}
\e\label{Sab1}
S_{ab}^P(f) = \frac{h_{c,P}^2}{12 \pi^2 f^3} \Gm^P_{ab}(f),
\q 
where $h_c^P(f)$ is the characteristic strain of the polarization mode $P$. In this work, we only search for the TT and ST polarization modes. The TT mode is the usual tensor transverse mode containing the ``$+$" and ``$\times$" polarizations, and the ST mode is the spin-0 breathing polarization. The overlap functions $\Gm_{ab}(f)$ for the $\TT$ and $\ST$ polarizations can be approximated by \citep{Hellings:1983fr,2008ApJ...685.1304L}
\m\label{TTST}
\Gm^{\TT}_{ab}(f) &=& \hf (1+\dt_{ab})+ \frac{3}{2} k_{ab} \(\ln k_{ab}-\frac{1}{6}\), \\
\Gm^{\ST}_{ab}(f) &=& \frac{1}{8}\(3 + 4\dt_{ab} + \cos\xi_{ab}\),
\n 
where $\dt_{ab}$ is the Kronecker delta symbol, $\xi_{ab}$ is the angle between pulsars $a$ and $b$, and $k_{ab} \equiv (1-\cos\xi_{ab})/2$. The TT overlap function $\Gm^{\TT}_{ab}$ is also known as the Hellings \& Downs \citep{Hellings:1983fr} or quadrupolar correlations. Note that the TT and ST overlap functions are related to each other by $\Gm^{\TT}_{ab} = \Gm^{\ST}_{ab} + (3/2) k_{ab} \ln k_{ab}$. Here, we also consider a parameterized overlap function 
\e\label{alpha} 
 \Gm_{ab}(f) = \frac{1}{8}\(3 + 4\dt_{ab} + \cos\xi_{ab}\) + \frac{\alpha}{2} k_{ab} \ln k_{ab},
\q 
which reduces to $\Gm^{\ST}_{ab}$ when $\al=0$, and $\Gm^{\TT}_{ab}$ when $\al=3$.

We take the characteristic strain $h_c^P(f)$ to be a power-law form, which can originate from the SGWB produced by a population of inspiraling SMBHBs \citep{Jaffe:2002rt,Sesana:2008mz,Sesana:2008xk}. Assuming the binaries are in circular orbits and the orbital decay is dominated by the GW emission, the cross-power spectral density can be approximately estimated by \citep{Cornish:2017oic}
\e\label{Sab2}
    S_{ab}^P(f) = \Gm^P_{ab} \frac{\A{P}^2}{12\pi^2}   \(\frac{f}{\fyr}\)^{-\gamma_P} \fyr^{-3},
\q
where $\A{P}$ is the GW amplitude of the polarization mode $P$, and $\fyr = 1/\mathrm{year}$. The power-law index $\gamma_P$ is $13/3$ for the TT polarization, and $5$ for the ST polarization. The dimensionless GW energy density parameter per logarithm frequency normalized by the critical energy density for the polarization mode $P$ is related to $\A{P}$ by \citep{Thrane:2013oya}
\e 
    \ogw^{P}(f) = \frac{2\pi^2}{3 H_0^2} f^2 h_{c,P}^2 = \frac{2\pi^2\fyr^2}{3 H_0^2} \A{P}^2 \(\frac{f}{\fyr}\)^{5-\gamma_P},
\q 
where $H_0 = 67.4\, \km \sec^{-1} \Mpc^{-1}$ is the Hubble constant taken from Planck 2018 \citep{Aghanim:2018eyx}.

We now briefly describe the noise model used in our analyses. After subtracting the timing model from the TOAs, the timing residuals $\dt\bm{t}$ of each single pulsar are contributed from a number of sources by \citep[see \eg][]{Lentati:2016ygu}
\e\label{dt}
\dt\bm{t} = M \bm{\epsilon} + \bn{SN} + \bn{DM} + \bn{yrDM} + \bn{WN} + \bn{CP}.
\q
The first term $M \bm{\epsilon}$ accounts for the inaccuracies in the subtraction of timing model \citep[see \eg][]{Chamberlin:2014ria}, in which $M$ is the timing model design matrix obtained from \texttt{TEMPO2} \citep{Hobbs:2006cd,Edwards:2006zg} through \texttt{libstempo}\footnote{\url{https://vallis.github.io/libstempo}} interface, and $\bm{\epsilon}$ is a vector denoting small offsets for the parameters of timing model. The second term $\bn{SN}$ is the stochastic contribution from the red spin noise (SN) intrinsic to each pulsar and is modeled by a power law with $30$ frequency components. The third term $\bn{DM}$ denotes the stochastic contribution from the DM noise which is also modeled by a power law with $30$ frequency components. Unlike the SN, the DM noise is dependent upon the radio frequency whose information is added to the Fourier basis components. The fourth term $\bn{yrDM}$ is the stochastic contribution caused by annual DM variation described by a deterministic yearly sinusoid \citep[see \eg][]{Lentati:2016ygu}. The fifth term $\bn{WN}$ represents the stochastic contribution due to the white noise (WN), including a scale parameter on the TOA uncertainties (EFAC), an added variance (EQUAD), and a per-epoch variance (ECORR) for each backend/receiver system \citep[see \eg][]{Arzoumanian:2015liz}. We include separate EFACs and EQUADs for all the backend/receiver-dependent PTA data sets, and separate ECORRs for the backend/receiver-dependent NANOGrav data sets. The last term $\bn{CP}$ is the stochastic contribution due to the common-spectrum process (such as an SGWB) with the cross-power spectral density given by \Eq{Sab2}. In the analyses, we use $10$ frequency components roughly starting from $1.09\times 10^{-9}$Hz to $1.09\times 10^{-8}$Hz for the common-spectrum processes. For pulsar J1713+0747, we also include a chromatic exponential dip to model the sudden change in dispersion when the signal passes through the interstellar medium during propagation \citep{Lentati:2016ygu}.

\begin{table}[tbp!]
    \caption{\label{tab:BF}An interpretation of the Bayes factor in determining which model is favored, as given by \cite{BF}.}
    \begin{ruledtabular}
        \begin{tabular}{c c l}
            $\BF$ & $\ln\BF$ & Strength of evidence\\
            \hline
            $< 1$ & $< 0$ & Negative\\
            $1 - 3$ & $0 - 1$ & Not worth more than a bare mention\\
            $3 - 20$ & $1 - 3$ & Positive\\
            $20 - 150$ & $3 - 5$ & Strong\\
            $> 150$ & $> 5$ & Very strong\\
        \end{tabular}
    \end{ruledtabular}
\end{table}

We use the latest JPL solar system ephemeris (SSE) DE438 \citep{DE438} as the fiducial SSE as opposed to the DE436 \citep{DE436} that was used to create the IPTA DR2. To extract information from the data, we perform similar Bayesian parameter inferences based on the methodology in \cite{Arzoumanian:2018saf,Arzoumanian:2020vkk}. The model parameters and their prior distributions are summarized in \Table{tab:priors}. We first perform the parameter estimations for each single pulsar without including the stochastic contribution from the common-spectrum process (\textit{i.e.} the $\bn{CP}$ term in \Eq{dt}). To reduce the computational costs, we then fix the white noise parameters to their max likelihood values from single-pulsar analysis. We use \texttt{enterprise} \citep{enterprise} and \texttt{enterprise\_extension} \citep{enterprise_extensioins} software packages to calculate the likelihood and Bayes factors and use \texttt{PTMCMCSampler} \citep{justin_ellis_2017_1037579} package to do the Markov chain Monte Carlo sampling. Similar to \cite{Aggarwal:2018mgp,Arzoumanian:2020vkk}, we use draws from empirical distributions to sample the parameters from SN, DM noise, and annual DM variation, with the distributions based on the posteriors obtained from the single-pulsar Bayesian analysis, thus reducing the number of samples needed for the chains to burn in. 

\section{Results and discussion} 

Our analyses are mainly based on the Bayesian inference in which the Bayes factor
is used to quantify the model selection scores. The Bayes factor is defined as
 \e
\BF \equiv \frac{\rm{Pr}(\mathcal{D}|\mathcal{M}_2)}{\rm{Pr}(\mathcal{D}|\mathcal{M}_1)},
\q
where $\rm{Pr}(\mathcal{D}|\mathcal{M})$ denotes the probability that the data $\mathcal{D}$ are produced under the assumption of model $\mathcal{M}$. Model $\mathcal{M}_2$ is preferred if the $\BF$ is sufficiently large. An interpretation of the $\BF$ in model comparison given by \cite{BF} can be found in \Table{tab:BF}.

\begin{figure}[htbp!]
    \centering
    \includegraphics[width=\linewidth]{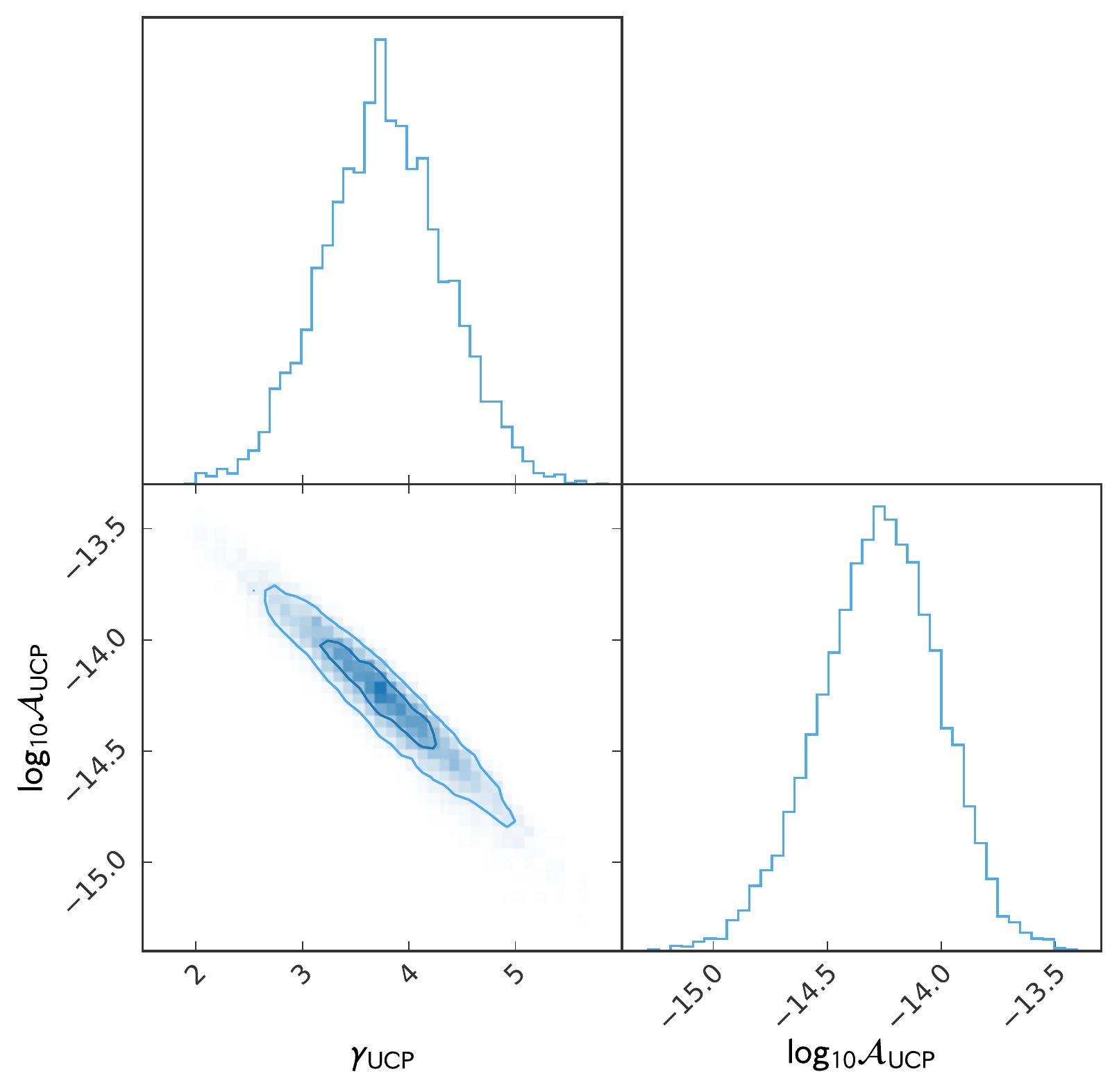}
    \caption{\label{post_UCP_A_gamma} 
        One and two-dimensional marginalized posteriors of the amplitude $\A{\mathrm{UCP}}$ and the power-index $\gamma_\mathrm{UCP}$ obtained from the UCP model with $\gamma_\mathrm{UCP}$ allowed to vary. We show both the $1 \sigma$ and $2 \sigma$ contours in the two-dimensional plot.}
\end{figure}

In \Table{bayes}, we summarize the Bayes factors of various models. The $\ln \BF$ of the UCP model with $\gamma_{\UCP}=13/3$ versus the noise only (NO) model without any common-spectrum process is $10.5$, indicating significant Bayesian evidence for a common-spectrum process in the IPTA DR2. When allowing the power-index to vary, the $\gamma_{\UCP}$ shows a relatively broad distribution as can be seen in \Fig{post_UCP_A_gamma}. The median value and $90\%$ equal-tailed credible intervals are $\log_{10} \A{\UCP}=-14.26^{+0.41}_{-0.47}$ and $\gamma_{\UCP}=3.76^{+0.96}_{-0.95}$.

\begin{table}[tbp!]
    \begin{center}
        \begin{tabular}{c|c|c|c}
            \hline\hline
            UCP vs. NO & TT vs. UCP & ST vs. UCP & ST+TT vs. UCP\\
            \hline
            $10.5(9)$ & $2.53(3)$ & $3.39(4)$ & $3.63(4)$\\
            \hline
        \end{tabular}
    \end{center}  
    \caption{\label{bayes}The $\ln \BF$ between pairs of models. The digit in the parentheses represents the uncertainty on the last quoted digit.}
\end{table}

\begin{figure}[tbp!]
    \centering
    \includegraphics[width=\linewidth]{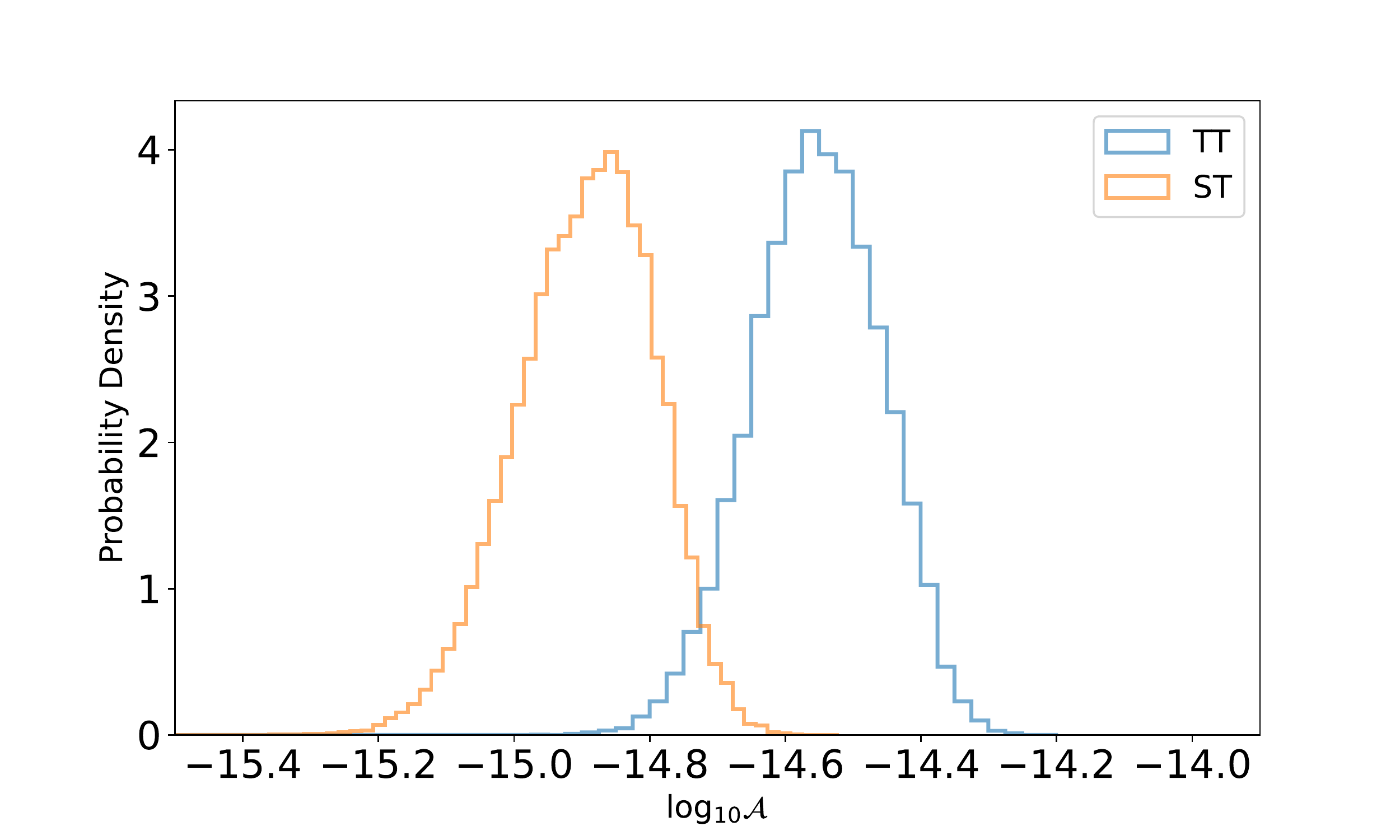}
    \caption{\label{post_TT_ST} Marginalized posteriors of $\A{\TT}$ and $\A{\ST}$ parameters obtained from the TT and ST models, respectively.}
\end{figure}

From \Table{bayes}, the $\ln \BF$ of the ST model versus the UCP (with $\gamma_{\UCP}=13/3$) model is $3.39$, indicating ``strong" Bayesian evidence of the ST correlations in IPTA DR2 according to \Table{bayes}. We obtain the amplitude as $\A{\ST} = 1.29^{+0.51}_{-0.44} \times 10^{-15}$ or equivalently $\ogw^\ST = 2.31^{+2.19}_{-1.30} \times 10^{-9}$, at frequency of 1/year. This result is consistent with the one reported in \cite{Chen:2021wdo} where $\Omega_{\mathrm{GW}}^{\mathrm{ST}}=1.54^{+1.20}_{-0.71}\times 10^{-9}$ from the NANOGrav 12.5-yr data set. On the other hand, the $\ln \BF$ of the TT model versus the UCP (with $\gamma_{\UCP}=13/3$) model is $2.53$, indicating ``positive" Bayesian evidence of the TT correlations in IPTA DR2. As the evidence for the TT correlations is insignificant, we do not report the median value of the $\A{\TT}$ parameter here. Both the posteriors of the $\A{\TT}$ and $\A{\ST}$ parameters obtained respectively from the TT and ST models are shown in \Fig{post_TT_ST}.

\begin{figure}[htbp!]
    \centering
    \includegraphics[width=\linewidth]{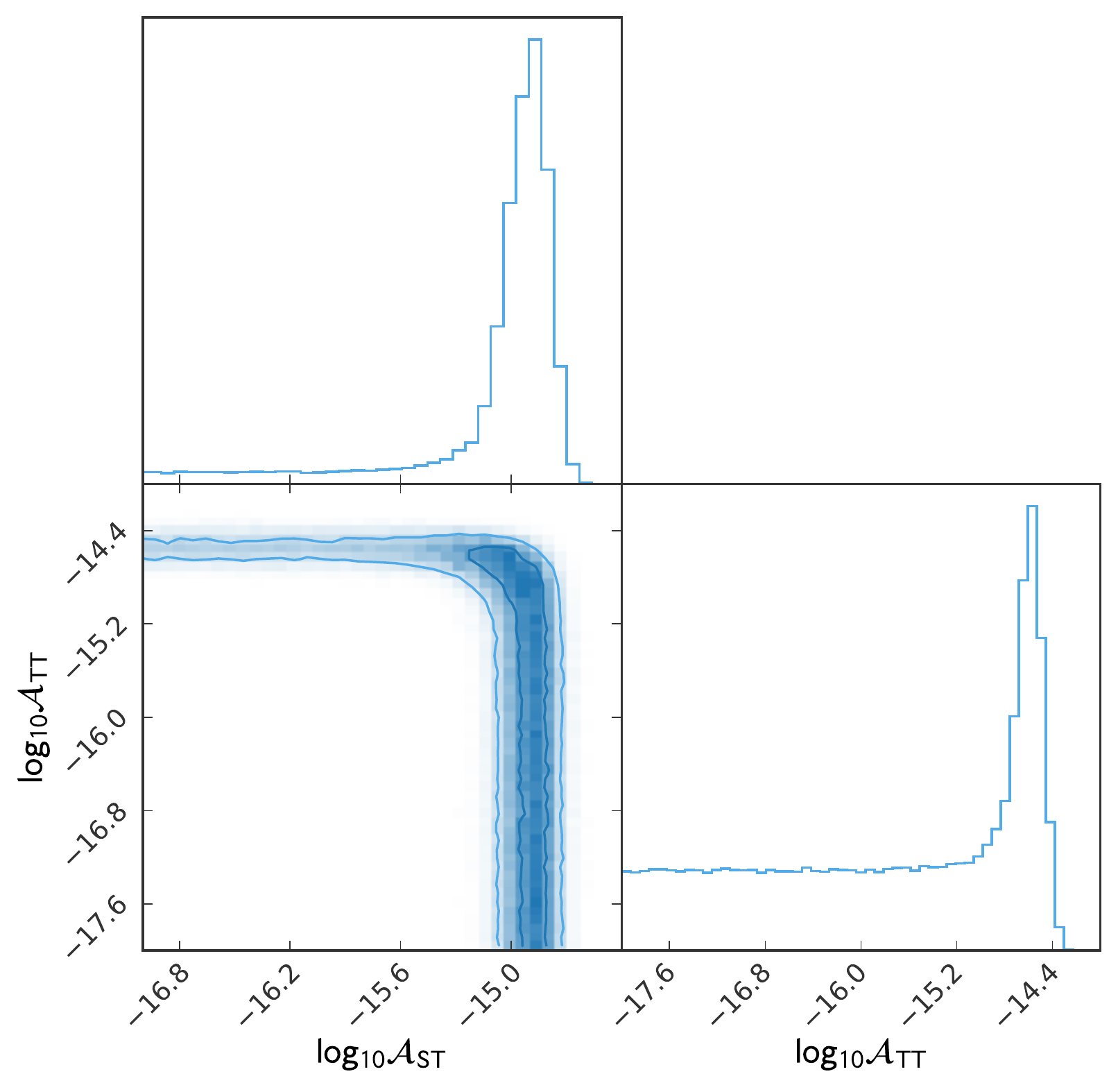}
    \caption{\label{post_ST+TT_A} One and two-dimensional marginalized posteriors of ST and TT amplitudes obtained from the ST+TT model. We show both the $1 \sigma$ and $2 \sigma$ contours in the two-dimensional plot.}
\end{figure}

Furthermore, we consider an ST+TT model where both the ST and TT spatial correlations are taken into account simultaneously. The $\ln \BF$ of the ST+TT model versus the UCP (with $\gamma_{\UCP}=13/3$) model is $3.63$, indicating no significant evidence for a second common-spectrum process with TT correlations on top of the ST-correlated process. The posterior distributions of the ST and TT amplitudes obtained from the ST+TT model are shown in \Fig{post_ST+TT_A}, with the amplitude of ST mode from this model is $\A{\ST} = 1.11^{+0.63}_{-1.10} \times 10^{-15}$ or equivalently $\ogw^\ST = 1.71^{+2.48}_{-1.70} \times 10^{-9}$, at frequency of 1/year. Note that this result is consistent with the one obtained from the ST model and that reported in \cite{Chen:2021wdo} where $\Omega_{\mathrm{GW}}^{\mathrm{ST}}=1.54^{+1.20}_{-0.71}\times 10^{-9}$ from the NANOGrav 12.5-yr data set.

\begin{figure}[tbp!]
    \centering
    \includegraphics[width=\linewidth]{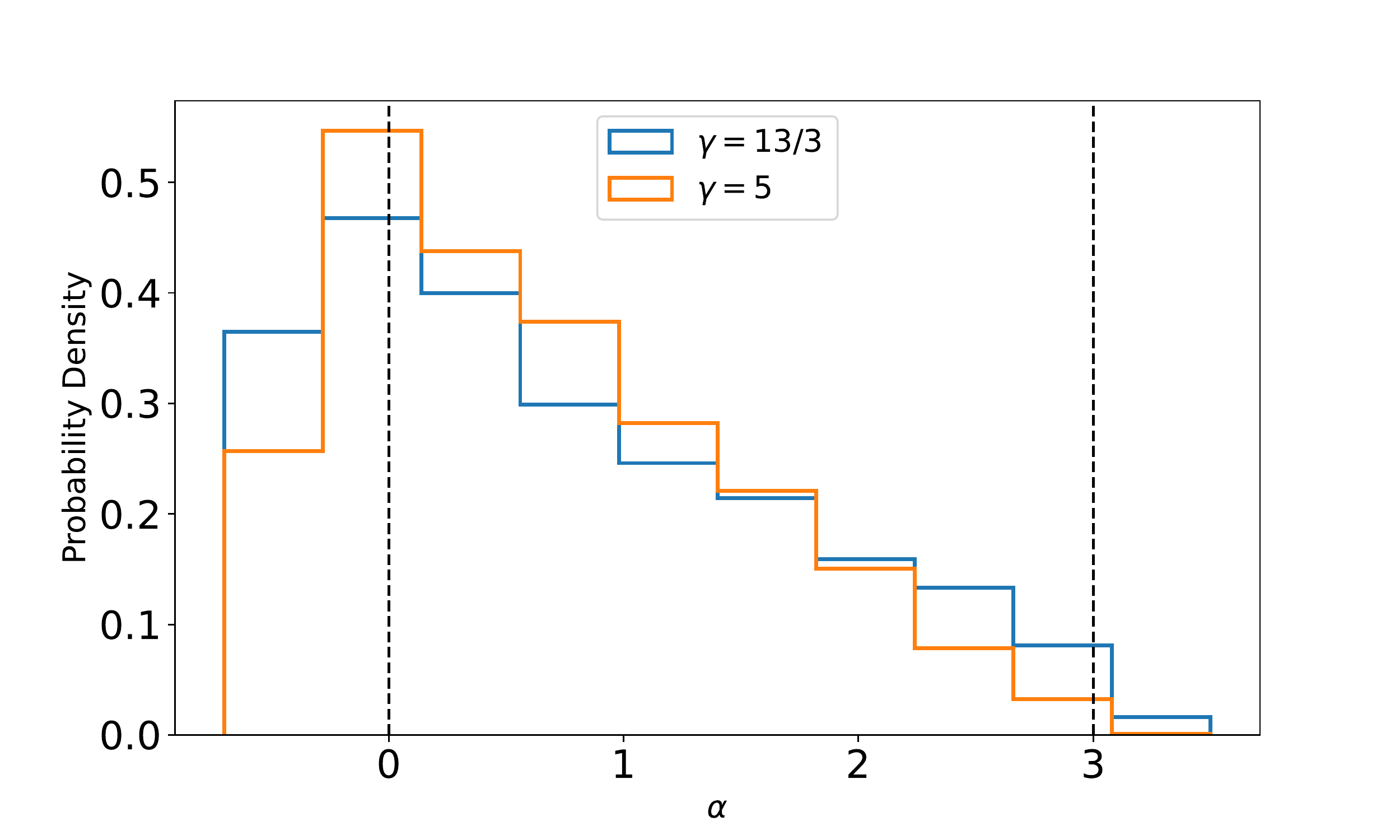}
    \caption{\label{post_alpha} Bayesian posteriors for the $\al$ parameter in the model with a parameterized overlap function of \Eq{alpha}. We consider the power-law SGWB spectrum with both $\gamma=13/3$ and $\gamma=5$. The two vertical dashed lines correspond to the TT ($\alpha=3$) and ST ($\alpha=0$) overlap functions, respectively.}
\end{figure}

To evaluate the potential deviation of the ST correlations in IPTA DR2, we also consider a model in which the overlap function is parameterized by the $\al$ parameter as \Eq{alpha}. Specifically, $\al=0$ corresponds to the ST correlations, while $\al=3$ corresponds to the TT correlations. The marginalized posteriors for the $\al$ parameter are shown in \Fig{post_alpha}, indicating the IPTA DR2 can be very well described by the ST correlations, but TT correlations are excluded by the $90\%$ credible regions.

Based on the above discussions, we conclude that there is strong Bayesian evidence for the ST correlations, but no significant evidence for the TT correlations in the IPTA DR2. Therefore, we place $95\%$ upper limits on the amplitude of TT polarization mode as $\A{\TT} \lesssim 3.95\times 10^{-15}$ and $\A{\TT} \lesssim 3.27\times 10^{-15}$, or equivalently, the $95\%$ upper limits for the energy density parameter per logarithm frequency are $\Omega_{\mathrm{GW}}^{\mathrm{TT}} \lesssim 2.16\times 10^{-8}$ and $\Omega_{\mathrm{GW}}^{\mathrm{TT}} \lesssim 1.48\times 10^{-8}$, from the TT and ST+TT models, respectively.

\begin{acknowledgments}
We acknowledge the use of HPC Cluster of ITP-CAS and HPC Cluster of Tianhe II in National Supercomputing Center in Guangzhou. This work is supported by the National Key Research and Development Program of China Grant No.2020YFC2201502, grants from NSFC (grant No. 11975019, 11690021, 11991052, 12047503),  the Key Research Program of the Chinese Academy of Sciences (Grant NO. XDPB15), Key Research Program of Frontier Sciences, CAS, Grant NO. ZDBS-LY-7009, and the science research grants from the China Manned Space Project with NO. CMS-CSST-2021-B01.
\end{acknowledgments}
\newpage
\bibliography{ref}
\bibliographystyle{aasjournal}

\end{document}